\documentclass[twocolumn,secnumarabic,amssymb, nobibnotes, 10pt, aps, prl, superscriptaddress]{revtex4-2}
\usepackage{graphicx}
\usepackage{amsfonts}
\usepackage{amssymb}
\usepackage{amsmath}
\usepackage{psfrag}
\usepackage{natbib}
\usepackage{mathtools}
\usepackage[normalem]{ulem}
\usepackage{subfigure,float}
\usepackage[colorlinks=true, linkcolor=blue, urlcolor=blue, citecolor=black]{hyperref}

\begin{document}

\title{Reversal of Solvent Migration in Poroelastic Folds}

\author{Mees M. Flapper}
\affiliation{Physics of Fluids Group, Faculty of Science and Technology,
Mesa+ Institute, University of Twente, 7500 AE Enschede, The Netherlands}
\author{Anupam Pandey}
\affiliation{Department of Mechanical \& Aerospace Engineering and BioInspired Institute,
Syracuse University, Syracuse, NY 13244, USA}
\author{Stefan Karpitschka}
\affiliation{Max Planck Institute for Dynamics and Self-Organization, 37077 G\"ottingen, Germany}
\author{Jacco H. Snoeijer}
\affiliation{Physics of Fluids Group, Faculty of Science and Technology,
Mesa+ Institute, University of Twente, 7500 AE Enschede, The Netherlands}

\begin{abstract}
Polymer networks and biological tissues are often swollen by a solvent, such that their properties emerge from a coupling between swelling and elastic stress. This poroelastic coupling becomes particularly intricate in wetting, adhesion, and creasing, for which sharp folds appear that can even lead to phase separation. Here we resolve the singular nature of poroelastic surface folds and determine the solvent distribution in the vicinity of the fold-tip. Surprisingly, two opposite scenarios emerge depending on the angle of the fold. In obtuse folds such as creases, it is found that the solvent is completely expelled near the crease-tip, according to a nontrivial spatial distribution. For wetting ridges with acute fold angles, the solvent migration is reversed as compared to creasing, and the degree of swelling is maximal at the fold-tip. We discuss how our poroelastic fold-analysis offers an explanation for phase separation, fracture and contact angle hysteresis. 
\end{abstract}

\maketitle

Polymer networks can absorb large amounts of solvents, driven by the entropy of mixing. The solvent migration in or out of an elastic network causes significant change in volume through swelling or shrinking~\cite{flory1953principles}.
This interplay between liquid transport and elasticity, known as poroelasticity, is intrinsic to a number of biophysical processes such as belb formation in cells~\cite{charras2005non}, skin maceration (hyperhydration) due to prolonged water exposure~\cite{cutting2002maceration}, and crucial for technological applications of soft materials such as `plasticizers' for softening `plastics'~\cite{wei2019plasticiser}, hydro- and organogels in mechanobiology~\cite{smith2018differentiation} or electronic encapsulants with self-healing properties~\cite{yang2013selfhealing}.

\begin{figure}[t]
    \centering
    \includegraphics[width=86mm]{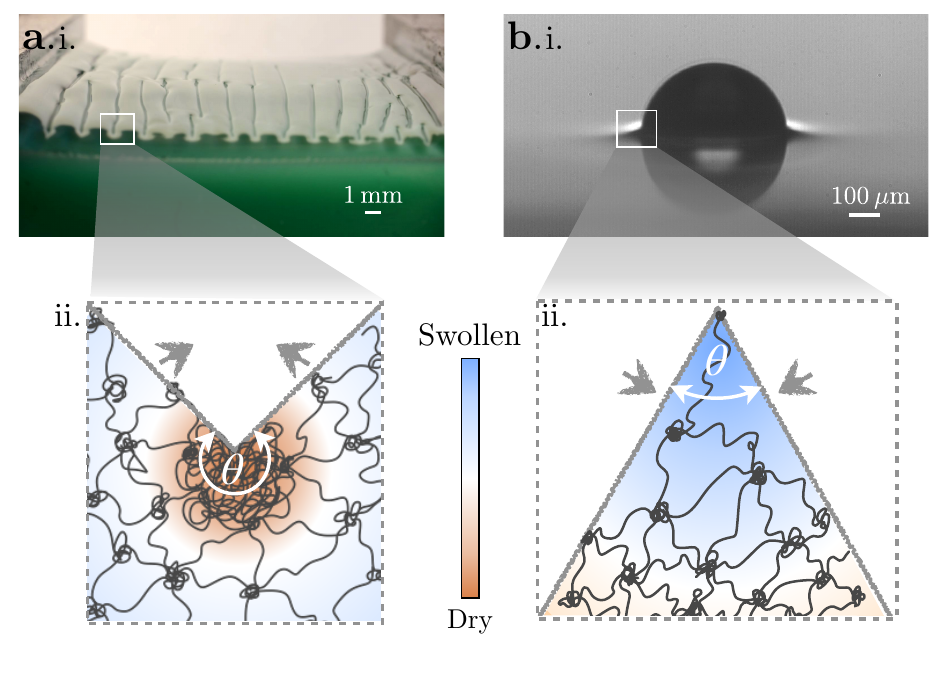}
    \caption{Poroelastic folds in (a) creasing, with fold angles $\theta>\pi$, and (b) wetting, with $\theta < \pi$, leading to opposite solvent distributions. (a.i) Swollen or compressed surfaces can spontaneously fold into self-contacting creases.  (a.ii) The obtuse fold gives rise to a blow-up of pore pressure, squeezing out the solvent and causing a completely dry fold tip. A self-contacting crease corresponds to $\theta=2\pi$. (b.i) A water drop on a soft PDMS gel, oblique side view. (b.ii) The acute fold angle at the wetting ridge causes a negative pore pressure, aspirating the solvent toward maximum swelling at the tip.}
           \label{fig:sketch}
\end{figure}

The surfaces of soft poroelastic matter can exhibit sharp folds that can be either acute or obtuse. 
Obtuse poroelastic folds are found in creases, when swollen polymer networks spontaneously develop morphologies~\cite{dervaux2012review, trujillo2008hydrogels, bertrand2016dynamics} where a surface folds onto itself [Fig.~\ref{fig:sketch}(a)]. These creases mimic the growth-induced gyrification of mammalian brains~\cite{mota2015cortical, tallinen2014gyrification, tallinen2016cortical} and tumors~\cite{dervaux2011tumors}. 
Folds with acute angles are observed in wetting~\cite{park2014visualization, jerison2011deformation,style2013universal,Andreotti:ARFM2020} and adhesion~\cite{style2013surface,chakrabarti2018elastowetting}, where the substrate is pinched into an acute ridge of well-defined opening angle [Fig.~\ref{fig:sketch}(b)].  
Both types of folds are expected to create a divergence of elastic stress~\cite{pandey2020singular}, but the implication of the fold geometry on the solvent distribution near the tip has barely been addressed~\cite{zhao2018growth}.
Yet, the singular stress-solvent interaction
will turn out to be central to unexplained observations such as the expulsion of solvent or un-crosslinked chains~\cite{cai2021fluid, jensen2015wetting}, wetting induced fracture~\cite{bostwick2013capillary, daniels2007instabilities}, and hysteric phenomena in both wetting and creasing~\cite{van2018dynamic,van2021pinning}.

In this Letter we analyze the poroelastic fold singularity, resolving both the distribution of stress and the degree of swelling at equilibrium.
We observe a reversal of solvent migration between ``creasing" and ``wetting", by the mechanism described in Fig.~\ref{fig:sketch}.
For creasing, the fold stretches the surface angle from $\theta=\pi$ (initially flat) to $\theta = 2\pi$; the polymer stress induced by this angular stretching generates a high pore pressure near the tip that `squeezes' the solvent out of the fold. The opposite trend is observed for wetting, where the surface angle is compressed to $\theta < \pi$, leading to a maximal aspiration of solvent towards the tip. We highlight how these intricate solvent distributions shed new light on recent experiments. 

\begin{figure*}[t]
    \centering
    \includegraphics[width=1\textwidth]{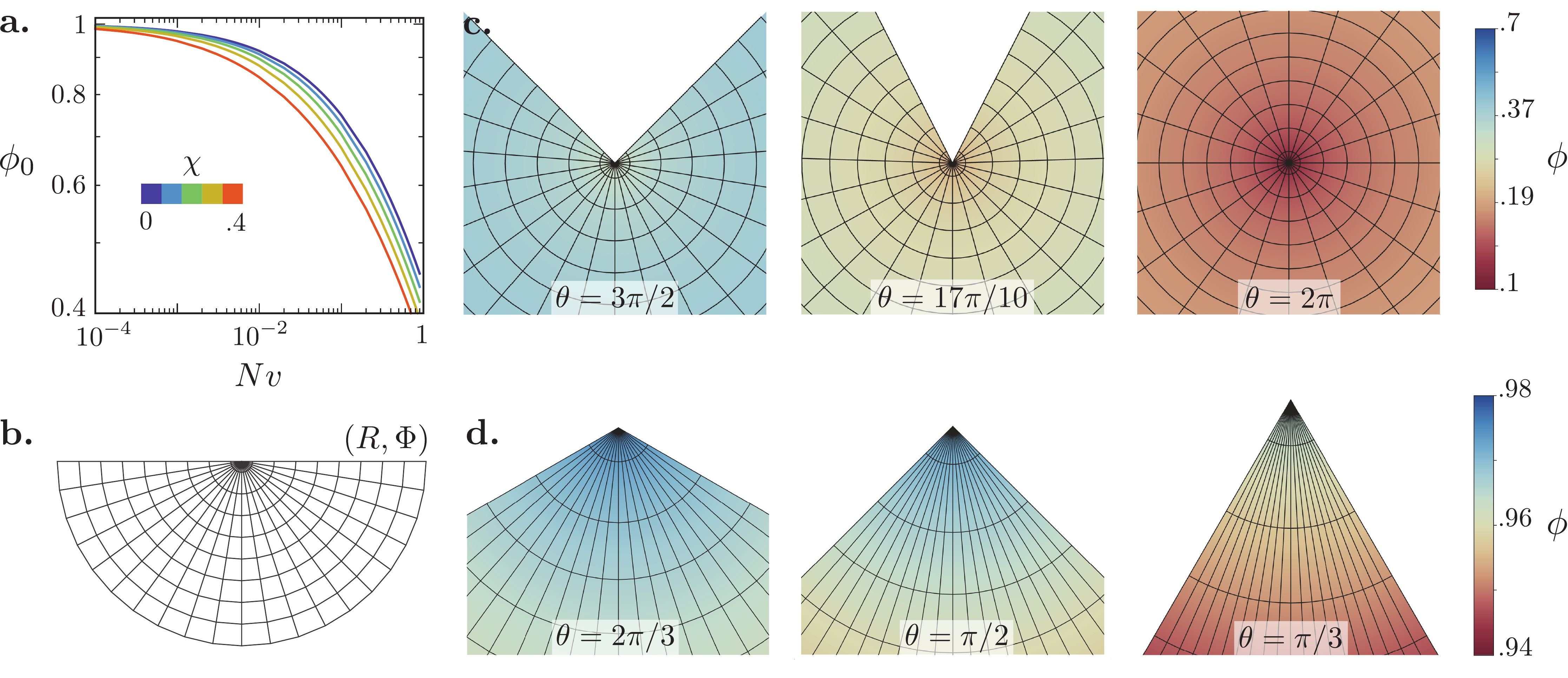}
    \caption{Solvent migration into or out of poroelastic folds. (a) Preswelling volume fraction $\phi_0$ as a function of $Nv$, for an unfolded substrate. (b)  A polar grid $(R,\Phi)$ characterises the material points in the unfolded substrate. (c,d) Various folds for $Nv=0.1$. The lines represent the deformed grid, folded to new positions $(r,\varphi)$. The color represents liquid volume fraction $\phi$, which is enhanced or reduced with respect to the preswelling $\phi_0 \sim 0.8$. (c) Creasing: upward folds for increasing fold angle $\theta$. A crease corresponds to $\theta=2\pi$. The liquid fraction within a crease is given by \eqref{phicrease}. The dark red spot in the center indicates the dry-out near the crease tip. (d) Wetting: downward folds with decreasing fold angles. In contrast to creasing, swelling is strongly enhanced near the tip (dark blue). In this case, $\phi$ is given by \eqref{phiridge}, plotted for $I_{\rm max}=100$.  }
           \label{fig:creasing}
\end{figure*}

\paragraph*{Poromechanics~---~}  

We start by considering a flat poroelastic substrate in the standard large-deformation framework \cite{hong2008theory}, which subsequently deforms into a planar, two-dimensional fold. The presence of the solvent results in swelling of the elastic polymer network, characterised by a swelling ratio $J$ that compares the swollen volume with that of the polymer network in the dry state. Assuming both the solvent and polymer molecules to be incompressible, the polymer volume fraction naturally becomes $1/J$, while the liquid volume fraction $\phi$ is given by the remainder

\begin{equation}\label{eq:phi}
\phi= 1 - 1/J.
\end{equation}
This expresses the relation between volume-change due to a stretching of the elastic network ($J$) and volume fraction of the solvent ($\phi$). Likewise, the free energy of a poroelastic material contains two contributions: a mixing energy that describes the interaction of the solvent with the polymers, and an elastic energy due to the stretching of the polymer network. Thus the resulting stress tensor $\boldsymbol{\sigma} =  - p \mathbf I + \boldsymbol{\sigma}_{\rm el} $, 
consists of a pore pressure (or osmotic pressure) $p$ and an elastic stress $\boldsymbol{\sigma}_{\rm el}$. According to Flory-Huggins theory~\cite{flory1942thermodynamics,huggins1942theory}, the pore pressure at equilibrium reads

\begin{equation}
p(\phi) =- \frac{kT}{v} \left[ \ln \phi  + (1-\phi) + (1-\phi)^2 \chi \right]+\frac{\mu}{v}, \label{eq:porepressure}
\end{equation}
with $v$ the volume per liquid molecule, $\chi$ the interaction parameter between solvent and network, and $\mu$ the solvent's reference chemical potential. The pore pressure is a decreasing function of $\phi$, and is always larger than the reference pressure $p(\phi=1)=\mu/v$ of the pure solvent.

The other component of the stress tensor, the elastic stress $\boldsymbol \sigma_{\rm el}$, is a function of the the Finger tensor $\mathbf B$, which characterises the stretching of the polymer network. The tensor is in general anisotropic, and the local volumetric swelling ratio follows from $J=\det (\mathbf B)^{1/2}$.
For long chain polymer networks, a commonly used constitutive relation between stress and strain is the so called ``neo-Hookean" model,
\begin{equation}\label{eq:sigmael}
\boldsymbol{\sigma}_{\rm el} (\mathbf B) = G (\mathbf B-\mathbf I) /J,
\end{equation}
where $G$ is the shear modulus of the polymer network. 

The degree of swelling prior to  folding depends on the ratio of the osmotic pressure scale ($kT/v$), to the network's elastic modulus ($G$).
The latter can be expressed as $G=NkT$, where $N$ represents the number of chains per unit volume \cite{hong2008theory}. 
With this, the problem is described by the dimensionless parameter $Nv=Gv/kT$.
The preswelling volume fraction $\phi_0$ of an isotropic medium, as described by the classical Flory-Rehner theory~\cite{flory1943swelling}, is then recovered from $\boldsymbol \sigma = -\mu/v \mathbf I$.
Typical values of $\phi_0$ can be inferred from Fig~\ref{fig:creasing}(a). We note that different hydrogels approximately cover the range $Nv = 10^{-4}\dots 10^{-1}$ \cite{hong2008theory}, while a polymer network swollen with un-crosslinked chains has $Nv \sim 1$.

\paragraph*{Fold geometry \& mechanics~---~}

The central aim is to resolve the stress singularity, which is manifestly anisotropic, and the resulting solvent distribution inside a fold. 
In polar coordinates, the most general, azimuthally symmetric fold is described by the mapping

\begin{equation}\label{eq:comprmap}
\varphi = b \Phi, \quad r/\lambda(r) = R,
\end{equation} 
where $(R, \Phi)$ characterizes the material points of the substrate in the undeformed state [cf. Fig.~\ref{fig:creasing}(b)]. The first equation expresses how an azimuthal angle $\Phi$ in the reference state (prior to folding) is deformed into a new angle $\varphi$. We assume that this azimuthal deformation is uniform and defined by a factor $b=\theta/\pi$, where $\theta$ is the global angle of the fold. The response in the radial direction is captured by the stretch, $\lambda(r)$, mapping the radial position of a point originally at $R$ to the new position $r$. For incompressible media, the conservation of volume dictates that $\lambda=b^{-1/2}$~\cite{singh1965note}. However, poroelastic networks change their volume due to swelling and $\lambda(r)$ needs to be found self-consistently. The Finger tensor $\mathbf B$ corresponding to (\ref{eq:comprmap}) reads (cf. Supplementary Information~\cite{SI})

\begin{equation}\label{eq:B}
\mathbf B =
\begin{pmatrix}
\left( \frac{\lambda}{1 - r\lambda'/\lambda}\right)^2  
& 0 \\
0 & (b \lambda)^2  \\
\end{pmatrix} 
\simeq 
\begin{pmatrix}
\lambda^2   & 0 \\
0 & (b \lambda)^2  \\
\end{pmatrix} ,
\end{equation}
where in the last step we used that $|r\lambda'/\lambda| \ll 1$ in the vicinity of the tip~\cite{SI}. 
With this, $J= \det( \mathbf B)^{1/2} = b\lambda^2$, and the stress near the tip reads

\begin{eqnarray}
\sigma_{rr} &=& -\bar p +  G b^{-1}, \,\quad
\sigma_{\varphi\varphi}  = - \bar p +  G b. \quad\label{eq:phiphi}
\end{eqnarray} 
Here the isotropic part of $\boldsymbol{\sigma}_{\rm el}$ is absorbed within the modified pressure $\bar p=p+G/J$, which makes the stress near the fold tip an explicit function of the fold angle, via $b=\theta/\pi$.

The final step is to establish $\bar p$ from mechanical equilibrium, $\boldsymbol{\nabla}\cdot\boldsymbol{\sigma}=\mathbf 0$. Owing to the azimuthal symmetry, this simplifies to a radial force balance 

\begin{equation}
\frac{d\sigma_{rr}}{dr} = \frac{1}{r}\left(\sigma_{\varphi \varphi} - \sigma_{rr} \right).
\label{eq:radialbalance}
\end{equation}
Integrating using (\ref{eq:phiphi}) gives the pressure inside the fold,

\begin{equation}\label{eq:logpressure}
\bar p(r) =\bar p_0 - G  \left(b - b^{-1} \right) \ln r/r_0.
\end{equation}
The integration constant is conveniently expressed as a length scale $r_0$, which represents the typical distance from the tip where we recover the pressure $\bar p_0$ in the bulk of the preswollen medium~\footnote{The value of $r_0$ cannot be determined from the local corner analysis, but is inherited from matching to an outer geometry \cite{van2021pinning}.}.

\paragraph*{Creasing.~---~} 
Having solved for the stress, we are in a position to reveal the solvent distribution inside poroelastic folds. 
It is crucial to distinguish creasing ($\theta = 2\pi$) from wetting ($\theta < \pi$), since this changes the sign of the pore pressure. Specifically, for $b = \theta/\pi > 1$, the logarithmic divergence in \eqref{eq:logpressure} is positive, which is consistent with the pore pressure \eqref{eq:porepressure}. Namely, in the limit of $\phi\to 0$, the pore pressure gives $\bar p(\phi)\simeq -(kT/v) \ln \phi$, so that we find the solvent distribution 

\begin{equation}
\phi = \left( r/r_0\right)^\beta, \quad \mathrm{with} \quad \beta = Nv \left(b - b^{-1} \right) >0.
\label{phicrease}
\end{equation}

A central finding is thus that creasing induces an algebraic decay of the solvent fraction toward the fold tip. The exponent $\beta$ is nontrivial, involving both the network properties $Nv$ and the fold angle $b=\theta/\pi$. Noting that $\beta >0$ for any angle $\theta > \pi$  implies that solvent is expelled from the tip. Figure~\ref{fig:creasing}(c) shows this solvent distribution inside the fold for $Nv=0.1$. Indeed, the volume fraction is everywhere below the preswelling bulk value ($\phi_0\sim 0.8$). Most dramatically, the dark zone at the center indicates the complete dry-out of the crease tip. 

\paragraph*{Wetting.~---~} 
Poroelastic wetting corners are fundamentally different in nature from creasing folds. Namely for $b<1$, mechanical equilibrium (\ref{eq:logpressure}) requires a diverging negative pore pressure, aspirating liquid toward the fold tip. However, the lowest possible pressure that can be achieved is that of the pure solvent, $\mu/v$, achieved for maximum swelling $\phi = 1$ or $J \to \infty$. So even an ``infinite swelling-ratio" does not lead to an equilibrium state for folds with $b<1$. This paradox was already anticipated from a small-deformation linear poroelastic calculation \cite{zhao2018growth}, but now it is clear that the problem persists in the large-deformation neo-Hookean framework. 

The paradox can be resolved by accounting for \emph{strain-stiffening} of the polymer network. For large swelling, the polymers stretch to a degree that they can no longer be considered Hookean springs. This effect can be accounted for by an effective elastic modulus $ \psi(I)G$, where $\psi(I)$ is an increasing function of the mean-stretch $I=\mathrm{tr}(\mathbf B)$ \cite{green1992theoretical}. Reanalysing (\ref{eq:radialbalance}) with $\boldsymbol \sigma_{\rm el} \simeq \psi G \mathbf B$, we find a rather different stress field

\begin{equation}\label{eq:psiscaling}
\sigma_{rr} =  (r_0/r)^{1-b^2} G b^{-1}, \quad \Rightarrow \quad \psi = (r_0/r)^{1-b^2}.
\end{equation}
For $b<1$, this gives a stress-singularity that is now carried by the strain-stiffening of the polymer network, rather than by the pore pressure. Specifically, the divergence of $\sigma_{rr}$ implies that the polymers become maximally stretched at the tip.

Yet, the degree of swelling indeed remains finite within the standard framework for strain-stiffening. For example in the Arruda-Boyce \cite{arruda1993three} or the Gent model \cite{gent1996new,horgan2015remarkable}, the function $\psi$ diverges when $I$ reaches its maximum extensibility $I_{\rm max}$, which indeed comes with a finite swelling ratio.  For both the Arruday-Boyce and Gent model, the stress (\ref{eq:psiscaling}) implies a volume fraction (cf. Suppl.~\cite{SI})

\begin{equation}
\phi \simeq 1 - \frac{b^{-1}+b}{I_{\rm max}}  \left[ 1 + \left(\frac{r}{r_0}\right)^{1-b^2}\right].
\label{phiridge}
\end{equation}
From these results it is clear that strain-stiffening, encoded via $I_{\rm max}$, offers a regular solution for $\theta<\pi$, with a finite $\phi$ at the ridge tip. 

Figure~\ref{fig:creasing}(d) illustrates the strongly enhanced swelling within wetting ridges. The colormap indicates that the solvent fraction is much larger than that of the preswollen medium $\phi_0$. From the circular gridlines, one also infers the very large stretching in the radial direction, which is bounded only by the finite extensibility of the polymers.
Comparing creasing and wetting [Figs.~\ref{fig:creasing}(c/d)], we indeed find a reversal of the solvent migration away from/towards the fold tip. 

\paragraph*{Implications for polymer networks.~---~} 

What are the implications of these continuum singularities for experimental systems? For creasing, the singularity of pressure is relatively weak (logarithmic) and compressive in nature. In practice we estimate the value of the logarithm in (\ref{eq:logpressure}) in the range 1-10, when, \emph{e.g.}, taking $r$ to vary from molecular scales to the crease length $r_0$ (typically $100 \, \mu$m). Depending on the material, such compressive stresses can typically be accommodated without any material failure. Thus, these deformations are expected to be reversible, corroborated by experiments showing the reversibility of creasing without material failure~\cite{yoon2010swelling,cai2012elastomer,chen2012surface}. 

By contrast, the singularity of wetting requires further interpretation. The divergence of stress for a strain-stiffening model is much stronger (algebraic) and tensile in nature, and even requires the polymers to become fully stretched near the fold tip. Even when considering a moderate range of $r/r_0$, the predicted stress near the fold-tip exceeds $G$ by orders of magnitude.
Based on this prediction, we thus expect irreversible material failure to occur close to the tip.

\begin{figure}[t]
    \centering
    \includegraphics[width=0.48\textwidth]{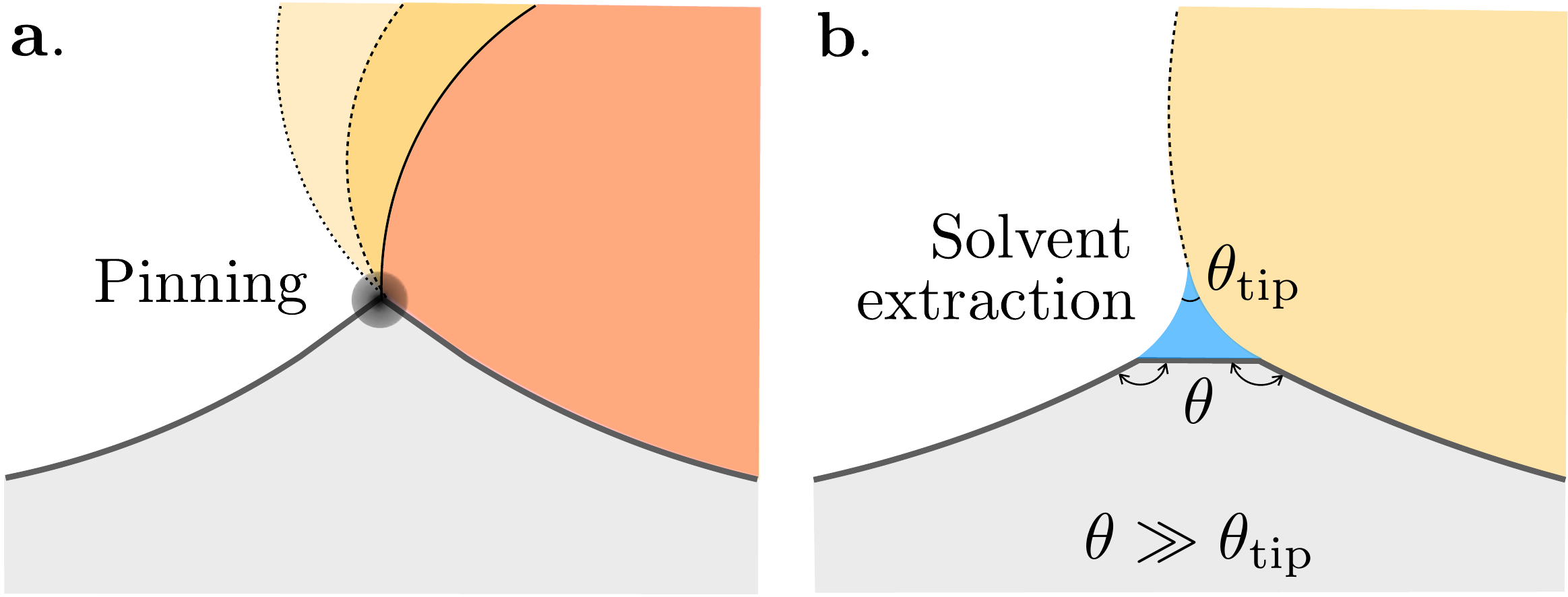}
    \caption{Sketch of poroelastic wetting scenarios. (a) The stress singularity predicted by our theory can lead to a local, irreversible fracture of the polymer network. This can act as a site for contact line pinning (observed in hydrogels \cite{kim2021measuring}). (b) Alternatively, the extraction of a liquid phase (observed for macromolecular solvents \cite{jensen2015wetting,cai2021fluid}) prevents sharp poroelastic folds and pinning due to irreversible damage is avoided. 
    }
           \label{fig:interpretation}
\end{figure}

On hydrogels, such wetting-induced fracture has indeed been observed experimentally~\cite{daniels2007instabilities,bostwick2013capillary}. The appearance of local fracture at the fold tip, predicted from our analysis, thus offers an explanation for strong contact line pinning observed recently~\cite{kim2021measuring} on hydrogels [Fig.~\ref{fig:interpretation}(a)]. 
Interestingly, however, wetting experiments on poroelastic cross-linked polydimethyl-siloxane (PDMS), which is ``swollen" by uncross-linked polymer chains, do not exhibit any significant contact angle hysteresis -- suggesting that pinning defects are absent. The key difference with hydrogels is the polymeric i.e., macromolecular nature of the swelling fluid. This significantly reduces the mixing entropy as compared to hydrogels [$N v \sim 1$ instead of $\sim10^{-4}$, see Fig.~\ref{fig:creasing}(a)] and enables the extraction of the fluid phase into a small meniscus with negative Laplace pressure.
This scenario is sketched in Fig.~\ref{fig:interpretation}(b): the formation of a sharp poroelastic fold is avoided by the extraction of solvent, which dramatically reduces the magnitude of stress.
This effect has been observed experimentally in the context of adhesion~\cite{jensen2015wetting} and wetting~\cite{cai2021fluid}. The precise conditions for solvent extraction remain to be identified, but it is evident that the geometric focussing of poroelastic stress plays a central role. 

\paragraph*{Outlook.~---~} 
We have demonstrated that poroelastic folds generate intricate solvent distributions that are fundamentally different between wetting and creasing, and which are governed by nontrivial exponents. Our analysis offers an explanation for wetting-induced pinning and fracture, and opens a new route to poroelasicity under extreme deformations. For example, while our analysis is restricted to equilibrium, it is clear that the divergence of stress will persist in the transient diffusion of solvent. On shorter time scales, solvent depletion renders the the crease tip a singular point, breaking the spatial invariance of the surface. This can cause pinning and hence might explain the annealable scars left behind by unfolded creases~\cite{van2021pinning,yoon2010swelling}. 
More generally, our findings show that the continuum framework of poroelasticity can break down near singular points, which must be addressed in future experiments or molecular simulations.

\begin{acknowledgments}
The authors thank J. Eggers for  discussions. JHS acknowledges financial support from NWO Vici (No. 680-47-63), and SK/JHS acknowledge support from the University of Twente-Max Planck Center for Complex Fluid Dynamics, and funding from the German research foundation (DFG, Project No. KA4747/2-1). AP acknowledges start up funding from Syracuse University.
\end{acknowledgments}


%

\end{document}